# CEREBELLAR PURKINJE CELL LOSS IN HETEROZYGOUS $RORA^{+/-}$ MICE: A LONGITUDINAL STUDY

MOHAMED DOULAZMI[1]*[†], FRANCESCA CAPONE[2][†], FLORENCE FREDERIC[†], JOËLLE BAKOUCHE[†], YOLANDE LEMAIGRE-DUBREUIL[†] AND JEAN MARIANI[†][‡].

[†] Université Pierre et Marie Curie–Paris6, Unité Mixte de Recherche (UMR) 7102–Neurobiologie des Processus Adaptatifs (NPA); Centre National de la Recherche Scientifique (CNRS), UMR 7102–NPA, Paris, F-75005 France.
[‡] APHP, Hôpital Charles Foix, UEF, Ivry sur Seine, F-94200.

1 & 2 contributed equally to this work

*Corresponding author: Tel.: (33) 1.44.27.35.08; Fax: (33) 1.44.27.22.80; E-mail: mohamed.doulazmi@snv.jussieu.fr .

**Summary:**

The staggerer (sg/sg) mutation is a spontaneous deletion in the Rora gene that prevents the translation of the ligand-binding domain (LBD), leading to the loss of RORα activity. The homozygous $Rora^{sg/sg}$ mutant mouse, whose most obvious phenotype is ataxia associated with cerebellar degeneration, also displays a variety of other phenotypes. The heterozygous $Rora^{+/sg}$ is able to develop a cerebellum which is qualitatively normal but with advancing age suffers a significant loss of cerebellar neuronal cells. A truncated protein synthesized by the mutated allele may play a role, both in $Rora^{sg/sg}$ and $Rora^{+/sg}$. To determine the effects during life span of true haplo-insufficiency of the RORα protein, derived from the invalidation of the gene, we compared the evolution of Purkinje cell numbers in heterozygous Rora knock-out males ($Rora^{+/-}$) and in their wild-type counterparts from 1 to 24 months of age. We also compared the evolution of Purkinje cell numbers in $Rora^{+/-}$ and $Rora^{+/sg}$ males from 1 to 9 months.

The main finding is that in $Rora^{+/-}$ mice, when only a half dose of protein is synthesized, the deficit was already established at 1 month and did not change during life span. Thus, the effects of aging on PC number were apparent much earlier in $Rora^{+/-}$ than in $Rora^{+/sg}$, although at 24 months of age the deficit was similar.

**INTRODUCTION**

Many naturally occurring mutations, described in mice, have been shown to affect the structure and/or the physiology of the cerebellum. The *staggerer* mutation, located on chromosome 9, consists of a deletion of an exon encoding part of the ligand binding domain within the Retinoic acid receptor-related Orphan Receptor (Rora) gene (Hamilton et al., 1996; Matysiak-Scholze and Nehls, 1997), leading to the loss of RORα activity (Hamilton et al., 1996) and to symptoms similar to those of RORα-null mice created by gene targeting (Doulazmi et al., 2001; Dussault et al., 1998). RORα has long been considered as a constitutive activator of transcription in the absence of an exogenous ligand but recent structural experiments have identified cholesterol or cholesterol sulfate as the natural ligand of RORα (Kallen et al., 2004; Kallen et al., 2002). However it is not clear if cholesterol is a "real" ligand rather than just structural cofactor and if changes in intracellular levels of cholesterol are capable of modulating the transcriptional activity of RORα in vivo. Low intracellular levels of cholesterol dramatically decrease the transcriptional activity of RORα in culture cells, but men and mice with inborn error of cholesterol biosynthesis are not ataxic (see review see Boukhtouche et al., 2004; Gold et al., 2006).

The phenotype of Rora–deficient mice is complex since RORα is a widely expressed nuclear receptor (Dzhagalov et al., 2004; Giguere et al., 1994; Matysiak-Scholze and Nehls, 1997) and because of both the direct and indirect effects of the mutation in the RORα gene (for review see Boukhtouche et al., 2004; Gold et al., 2006; Jarvis et al., 2002). The most obvious phenotype of homozygous Rora knockout mice ($Rora^{-/-}$) and natural RORα deficient *staggerer* ($Rora^{sg/sg}$) mice is an ataxic gait, associated with a massive cerebellar degeneration due to a defect in Purkinje cell development (Herrup et al., 1996).

The role of RORα expands beyond the cerebellum. For instance, RORα, expressed in the suprachiasmatic nucleus of the hypothalamus, is involved in the control of circadian rhythms through the transcriptional regulation of BmalI, itself a transcriptional activator that is necessary for core oscillator function (Akashi and Takumi, 2005; Emery and Reppert, 2004; Guillaumond et al., 2005; Sato et al., 2004; Ueda et al., 2005). Mutant mice defective in Rora (s*taggerer*) have altered circadian feeding and locomotor activity patterns (Akashi and Takumi, 2005; Sato et al., 2004) and altered circadian secretion of corticosterone (Frederic et al 2006).

Moreover, outside the central nervous system, ROR$\alpha$ is implicated in osteogenic (Meyer et al., 2000) and myogenic (Lau et al., 1999) differentiation and provides protection against chronic inflammation and age-related degenerative processes including osteoporosis and atherosclerosis (for review see Boukhtouche et al., 2004; Gold et al., 2006; Jarvis et al., 2002)

Recent studies have aimed at identifying ROR$\alpha$ target genes in different cell types and organs. These genes cover a wide range of functions. ROR$\alpha$ controls the transcription of the genes encoding the Purkinje cell protein-2 and sonic hedgehog, proteins that are crucially involved in the development of the central nervous system (Gold et al., 2003; Matsui, 1997). The protooncogene N-*myc* appears also to be a ROR$\alpha$ target gene (Dussault and Giguere, 1997; Gold et al., 2003). N-*myc* is essential for normal embryonic organogenesis, and its oncogenic activity is associated with tumors of neuroendocrine and embryonic origins. In man, ROR$\alpha$ has recently been identified as a very large fragile site gene that is inactivated in multiple tumors, and the transcriptional level of ROR$\alpha$ is down regulated in breast, ovary and prostate cancer (Zhu et al., 2006). The overexpression of the human ROR$\alpha$1 isoform in mouse cortical neurons protect them from apoptosis induced by apoptotic stimuli such as β-amyloid peptide, c2-ceramide and $H_2O_2$. This protection is due to an increase of the expression of the antioxidant proteins gluthathione peroxidase 1 and peroxiredoxin 6, leading to decrease amounts of stress-induced reactive oxygen species (Boukhtouche et al., 2006b).

In the cerebellum, ROR$\alpha$ expression is limited to the Purkinje cells (PCs) and interneurons of the molecular layer (Becker-Andre et al., 1993; Giguere et al., 1995; Ino, 2004; Steinmayr et al., 1998). Mice lacking functional ROR$\alpha$ protein (Steinmayer et al.; 1998, Hamilton et al.; 1996) have a very atrophic cerebellum (Doulazmi et al., 2001; Sidman et al., 1962). Vogel and co-workers (Vogel et al., 2000) have suggested that Purkinje cells are generated normally in the *staggerer* mutant, but many may fail to differentiate and/or start to die following birth. The early cerebellar atrophy is related to a massive reduction in the number of PCs (reduced by about 80%, Doulazmi et al., 2001) and study of chimeras has shown that this defect is due to an intrinsic action of the gene in the PCs (Herrup and Mullen, 1979). The lack of Purkinje cells induces a near total loss of their afferents, the granule cells and inferior olivary neurons, during the first postnatal month (Herrup et al., 1996). Part of the granule cell loss is due to the action of RORα on one of its target gene in the cerebellum, sonic hedgehog, since its addition to $Rora^{sg/sg}$ organotypic cultures restores granule precursor proliferation (Gold et al., 2003).

RORα controls the early steps of Purkinje cells dendritic differentiation. In homozygous Rora knockout mice ($Rora^{-/-}$) and natural RORα deficient *staggerer* ($Rora^{sg/sg}$) the surviving PCs, not arranged in a monolayer, display immature features. PCs are in an embryonic shape: somata appear smaller than control (Landis and Sidman, 1978), while dendrites are rudimentary and stunted (Doulazmi et al., 2001). In organotypic cerebellar cultures Rora-deficient Purkinje cells do not progress beyond the embryonic bipolar shape. Their infection with Lenti-hRORα1 restores both RORα expression and the dendritic differentiation (Boukhtouche et al., 2006a).

The heterozygous $Rora^{+/sg}$ develops a cerebellum which is qualitatively normal, but with advancing age suffers a significant loss of cerebellar neuronal cells. A loss of the same neuronal categories as in the homozygous mutant has a later onset, progressing from a normal cell numbers at 3 months of age to a deficit of 25-30% of the PCs and granule cells and 40% of olivary neurons at 12 months of age (Doulazmi et al., 1999; Hadj-Sahraoui et al., 1997; Zanjani et al., 1992). The Purkinje cell loss occurs earlier in *staggerer* males than in females. In males, the Purkinje cell loss starts from 1 month and continues regularly up to 13 months. In females, Purkinje cell number remains stable up to 9 months of age, then decreases to the same number as males (Doulazmi et al., 1999).

The heterozygous *staggerer* does not necessarily represent a situation of true haplo-insufficiency of the RORα protein. Given the molecular nature of the mutation, we cannot preclude a role for a truncated protein synthesized by the mutated alleles, both in $Rora^{sg/sg}$ and $Rora^{+/sg}$ mice. To determine the effects during life span of true haplo-insufficiency of the RORα protein, derived from the invalidation of the gene, we compared the evolution of Purkinje cell numbers in heterozygous Rora knock-out mice ($Rora^{+/-}$) and in their wild-type counterparts from 1 to 24 months of age. We also compared the evolution of Purkinje cell numbers in $Rora^{+/-}$ and $Rora^{+/sg}$ males from 1 to 9 months.

# MATERIALS AND METHODS

## Animals

The Rora knock-out mouse, was generated in 1998 by Steinmayr and coworkers, on a 129 / Ola mouse (Steinmayr et al., 1998) and then derived on a C57BL/6J background. It was maintained as a colony in our animal facility. Heterozygous mice had been mated with C57BL/6J for 6 generations at the beginning of our study. Purkinje cells were counted on heterozygous $Rora^{+/-}$ males aged 1, 3, 9, 18 and 24 months. Controls were $Rora^{+/+}$ littermates of the same age and gender. $Rora^{+/-}$ females were kept for reproduction.

The $Rora^{sg/sg}$ mutant mouse is maintained on a C57BL/6J genetic background as a colony in our animal facility. Purkinje cells were counted on heterozygous $Rora^{+/sg}$ males aged 1, 3, and 9 months. Wild-type controls were littermates of the same age and gender.

We used 3-5 animals per group. All animal procedures were performed under the guidelines established by "Le Comité National d'Ethique pour le Scienes de la Vie et de la Santé."

## Genotype Analysis

Intercrossing of fertile ($Rora^{+/-}$) or ($Rora^{+/sg}$) mice produces homozygous mutant mice ($Rora^{-/-}$) or ($Rora^{sg/sg}$) easily recognizable by their ataxic gait from post natal day 14-15, and both wild-type and heterozygous ($Rora^{+/-}$) or ($Rora^{+/sg}$) with indistinguishable behavior.

Since both in the homozygous ($Rora^{-/-}$) and heterozygous ($Rora^{+/-}$) the missing gene has been replaced by a reporter gene (β-Gal) (Steinmayr et al., 1998) and since it is already known that the Rora gene is highly expressed in the skin, the diagnosis of $Rora^{+/-}$ mice was done by β-Gal staining of tail skin.

Sections of tail biopsies from putative $Rora^{+/+}$ and heterozygous $Rora^{+/-}$ were cut and fixed in glutaraldheyde (0.05%) for 5 min at room temperature, washed three times in phosphate-buffered saline (PBS; 0.1 M, pH 7.4) and incubated for 4 hours at 37°C with β-Gal substrate containing 4 mM MgCl2, 2 mM K4Fe(CN)6, 2 mM K3Fe(CN)6, and 0.4 mg/ml 5-bromo-4-chloro-3-indolyl β-D-galactoside (X-Gal). Sections were washed in PBS and observed under a dissecting microscope. In the $Rora^{+/-}$ β-Gal activity was revealed by the presence of blue cells in the hair follicles (Steinmayr et al., 1998). $Rora^{+/-}$ were kept for the experimental group and $Rora^{+/+}$ mice used as wild type controls.

Putative young wild type C57Bl/6J and heterozygous *staggerer* ($Rora^{+/sg}$) mice were genotyped by PCR. Genomic DNA was extracted from tail biopsies and amplified in two sets of reaction, one for each allele.

The *staggerer* allele primers were: 5'-CGTTTGGCAAACTCCACC-3' and 5'-GATTGAAAGCTGACTCGTTCC-3'.

The + allele primers were: 5'-TCTCCCTTCTCAGTCCTGACA-3' and 5'-TATATTCCACCACACGGCAA-3'. The amplified fragments (318 bp + and 450 bp sg) were detected by electrophoresis on agarose gel.

**Histology**

*Tissue preparation.*

After deep anesthesia, the animals were perfused transcardially with physiological saline followed by 4% paraformaldehyde in phosphate-buffered saline (PBS; 0.1 M, pH 7.4). After removal from the skull brains were postfixed overnight in fresh fixative.

In addition one animal from each genotype of each age group was perfused with physiological saline followed by 95% ethanol. The brain was dissected out of the skull and postfixed overnight in Clarke's fixative (3 volume of ethanol for 1 volume of acetic acid) to process for anti-28kd calcium binding protein (CaBP) immunocytochemistry for qualitative analysis, as previously described (Doulazmi et al., 2001).

All brains were dehydrated throughout a series of graded alcohols, embedded in paraffin and serially cut in 8 μm sagittal sections. Sections were stained with cresyl violet, counted, and photographed using a stereoscopic microscope (Nikon).

*Cell Counts*

PC counts were performed at 1, 3, 9, 18, 24 months with 3-4 animals per group. PCs were counted in every 40th section of the whole cerebellum as described previously (Doulazmi et al., 1999). In each section, each cell that was located in the PC layer, had a large soma and at least a portion of its nucleus in the section was counted. All counts were done on sections from coded animals by the same investigator. Duplicate counts of the same section did not differ by more than 3%. The number of Purkinje cells in each section was plotted as a function of the distance from the left paraflocculus. In each animal, counts were multiplied by forty to estimate the total number of cells in the cerebellum. These raw values were overestimates because cell nuclei can be split during sectioning, and thus may appear in more than one section; therefore, the corrected number of PCs was obtained by multiplying the raw values by the Hendry correction factor for each animal (Hendry, 1976). We chose to use this traditional correction factor for our cell counts instead of more recently developed stereological techniques (Andersen et al., 1992; Williams and Rakic, 1988), so that our results would be directly comparable with previously published PC counts including our own (Doulazmi et al., 1999; Hadj-Sahraoui et al., 1996; Hadj-Sahraoui et al., 1997; Herrup and Mullen, 1979). Moreover comparable results about the size of the PC population have been obtained using both techniques (Vogel et al., 1989).

The laterolateral extent of the cerebellar cortex was obtained by multiplying the thickness of the section by the total number of parasagittal sections containing PCs. The distances were expressed as a percentage of the total cerebellar width, and, at each age point, PC counts for homologous sagittal planes were averaged to compare the laterolateral distribution of the PC number in the heterozygous and wild type mice (See review in Herrup et al., 1996).

*Morphometry*

The areas of Purkinje cell somata were measured in sections from the vermis and the hemispheres. At each age, 50 cells per region were measured with the NIH image computer program.

**Statistics**

The differences between the mean number of PCs per cerebellum, mean laterolateral extents of cerebella, and Purkinje cell soma areas were assessed by a two way (genotype and age) analysis of variance (ANOVA) with post-hoc multiple comparison analysis using the Newman-Keuls test. At each age, the comparison of the laterolateral distribution of the averaged number of PCs per homologous sagittal plane was assessed by a two-way ANOVA (genotype and repeated measures on the sagittal plane).

**RESULTS**

**Comparative analysis of heterozygous *Rora* $^{+/-}$ versus *Rora* $^{+/+}$ males**

*Qualitative Analysis*

In heterozygous *Rora*$^{+/-}$ mice aged 1, 3, 9, 18 and 24 months, the cerebellar cortex retained its normal foliation and trilaminar organisation when compared to Rora$^{+/+}$ animals. There were no obvious changes in the folial pattern or cellular architecture. PCs remained easily identifiable by the location of their soma in a monolayer and normal morphology. However a slight decrease in the cross-sectional area of *Rora*$^{+/-}$ cerebella compared to *Rora*$^{+/+}$ was noticeable when sagittal sections were carefully examined (Figure. 1).

Anti-calbindin immunochemistry stains all parts of the Purkinje cell (soma, dendrites and axons). In both wild-type and heterozygous *Rora*$^{+/-}$, the PCs appeared evenly stained and no ectopic PCs were detected in the granular or the molecular layer (Figure 2A, 2B). In the heterozygous *Rora*$^{+/-}$, there was no obvious topographical repartition of the Purkinje cell loss. A higher magnification shows that, in both genotypes, the PCs were profusely branched and extended dendrites to the pial surface (Figure 2C and 2D).

*Purkinje cell loss*

The evolution of the PC population with age was different in *Rora*$^{+/+}$ and *Rora*$^{+/-}$ mice (Figure 3). There was a significant interaction between genotype and age (P < 0.05). In the *Rora*$^{+/+}$ the PC number remained stable through 1-18 months although there was a significant loss in 24-month-old mice (P < 0.001). Interestingly, no cell loss was observed in Rora$^{+/-}$ mice, the number of PCs remaining in fact stable during the entire time period considered (1 - 24 months). Nevertheless, in the *Rora*$^{+/-}$ the number of PCs was already significantly lower than in age-matched *Rora*$^{+/+}$ at 1 month of age (P < 0.001) and this difference persisted in 18-month-old heterozygous mice, with a mean deficit of 21%. By contrast, the difference in the number of PCs observed in *Rora*$^{+/+}$ and *Rora*$^{+/-}$ disappeared at 24 months when a loss of PCs was found in wild type animals.

*Regional distribution of the Purkinje cell loss*

The mean laterolateral extent of the cerebella within the different experimental groups ranged from 6790 to 7730 μm. There was no significant variation of the lateral extent between genotypes or ages. Figure 5 illustrates the averaged laterolateral repartition of the cell counts at selected ages in $Rora^{+/+}$ and $Rora^{+/-}$ mice. In $Rora^{+/-}$, the absence of interaction indicated that the cell loss occurred evenly throughout the laterolateral extent of the cerebellum. This deficit was also evenly distributed in 24-month-old $Rora^{+/+}$ mice.

## Comparative analysis of heterozygous $Rora^{+/-}$ versus $Rora^{+/sg}$ males

*Purkinje cell loss*

The evolution of the PC population with age was different in $Rora^{+/sg}$ and $Rora^{+/-}$ mice (Figure 5). The mean corrected number of PCs at different ages (1, 3 and 9 months) was significantly lower in $Rora^{+/-}$ than in $Rora^{+/sg}$ mice (p<0.001).

*Quantitative analysis of PC cell somatic size*

The somatic area of Purkinje cells was measured in the vermis and the hemispheric regions. There was no difference in size according to age. In both genotypes the cell bodies were slightly but significantly smaller in the hemispheres than in the vermis. The mean area of $Rora^{+/-}$ Purkinje cell bodies was 7.3 % smaller than $Rora^{+/sg}$ in the vermis and 3.8% smaller in the hemispheres (p<0.0001). This difference in area reflects a volumetric difference which would be more striking.

**DISCUSSION**

In the present study, the main finding is that, although at 24 months PC number became similar, the effects of aging on PC number were much more precocious in $Rora^{+/-}$ than in $Rora^{+/sg}$. The maximum of the PC loss as compared to wild types was obtained as early as 1 month of age in $Rora^{+/-}$ mice whereas this maximum loss was reached only after 13 months in $Rora^{+/sg}$ mice (Doulazmi et al., 1999). From these results, the phenotype of the $Rora^{+/-}$ appeared slightly more severe than that of the $Rora^{+/sg}$, as a $Rora^{+/sg}$ phenotype appears only with aging. This increase of severity is supported by the smaller size of the heterozygous $Rora^{+/-}$ Purkinje cell somata.

In heterozygous $Rora^{+/sg}$ males aged 3 and 9 months, the PC loss occurred predominantly in the intermediate region of the cerebellum and becomes evenly distributed only when the Purkinje cell loss has reached is maximum (Doulazmi et al 1999). By contrast, in $Rora^{+/-}$ males, the maximum of Purkinje cell loss is already reached at one 1 month of age and this loss is evenly distributed along the laterolateral axis. The difference of survival between $Rora^{+/-}$ and $Rora^{+/sg}$ Purkinje cells is probably not due to the expression of β-Gal in $Rora^{+/-}$ Purkinje cells. Indeed, in a previous study on homozygous $Rora^{-/-}$ mice (Doulazmi et al., 2001), we examined the surviving PCs for anti-CaBP staining and β-Gal activity on adjacent sections. We found a similar number of PCs positive for CaBP and for β-Gal activity. Since the expression of β-Gal does not affect the survival of $Rora^{-/-}$ PCs it seems unlikely that it may affect the survival of $Rora^{+/-}$ PCs.

We can only speculate on the factors involved in this difference. In $Rora^{-/-}$ and $Rora^{sg/sg}$ mice the phenotypes are identical: the loss of RORα protein at a critical stage of development is so deleterious that even if a slight difference in RORα protein existed between $Rora^{-/-}$ and $Rora^{sg/sg}$, its effect on the phenotype would be too small to be visible. By contrast, the phenotype is slightly more severe in $Rora^{+/-}$ than in $Rora^{+/sg}$ mice. The most likely explanation is that the $Rora^{+/sg}$ mouse is a hypomorphic mutant: the DNA binding domain may be synthesized as a truncated peptide which can not bind its ligand but could bind its target DNA. In $Rora^{+/sg}$ mice, a normal cerebellum is built and then a delayed and slow process of neuronal atrophy and retraction can occur: during this process many factors can be involved such as inflammation and hormonal modulations as we have already suggested (Doulazmi et al., 1999). In $Rora^{+/-}$ mice, when only a half dose of protein is synthesized, the deficit is more severe since it was already established at 1 month and did not change thereafter.

Consequently, the mutant Rorasg protein could compete for DNA binding sites or could sequester co-regulators present in limiting amounts, and therefore interfere with the normal function of other nuclear receptors or transcription factors active during Purkinje cell maturation.

Interestingly, in both homozygote and heterozygote mutant mice the phenotypes are comparable between the KO and the *staggerer* strains, with only a difference in the timing of the neuronal loss in the heterozygotes. This finding strongly suggests that the phenotype is due in both cases to a loss of function of the ROR$\alpha$ protein. This hypothesis has also been supported by the fact that overexpression of ROR$\alpha$ by transfection in culture has a neuroprotective effect on neurons (Boukhtouche et al, 2006b).

Further comparative molecular and cellular analysis of $Rora^{+/-}$ and $Rora^{+/sg}$ mice may reveal more subtle differences between the two mutant alleles and may provide additional insight into the molecular mechanisms of ROR$\alpha$ action in the development of the nervous system.

An increasing number of scientific articles report that the phenotype of a given single gene mutation in mice is modulated by the genetic background of the inbred strain in which the mutation is maintained (see review in Crusio, 2004). The knockout Rora mutation was generated in a C57BL/6 129/Ola hybrid strain. We studied the heterozygous $Rora^{+/-}$ after 6 generations of backcrossing with C557Bl/6J mice. We can not exclude the hypothesis that difference observed in PC number in the heterozygous $Rora^{+/sg}$ and the heterozygous $Rora^{+/-}$ mice might be due in part to the slight differences in their genetic background. However, we may suppose that with a higher degree of imbreeding the phenotype of $Rora^{+/-}$ mice would still be more severe.

**Acknowledgements**

We thank Dr. A Lohof for critical reading of the manuscript and P. Bouquet for excellent technical assistance in histology. F.C. was supported by fellowships from the Italian C.N.R. and the Fondazione Instituto Pasteur – Fondazione Cenci Bolognetti.

**Figure legends**:

Figure 1

Cerebella from wild type $Rora^{+/+}$ (A, B) and heterozygous $Rora^{+/-}$ (C, D) male mice at 1 (A, C) and 24 months (B, D) of age. Midsagittal sections stained with cresyl violet. The lobulation and the trilaminar organization that characterize the cerebellar cortex of wild type mice were also found in the heterozygous $Rora^{+/-}$ Scale bars = 1 mm.

Figure 2

Immunostaining with anti-calbindin antibody of cerebellar Purkinje cells on sagittal sections of cerebellum from 3-month-old $Rora^{+/-}$ (A) and $Rora^{+/+}$ (B) mice. The PCs appear evenly stained in both genotypes and no obvious topographical repartition of the cell loss is apparent in $Rora^{+/-}$. C and D: High magnification for anti-calbindin immunochemistry showing PC somata and dendritic arborisation. $Rora^{+/-}$ mouse (C), $Rora^{+/+}$ mouse (D). Scale bar = 500 μm (A and B) and 20 μm for (C and D).

Figure 3

Evolution of the Purkinje cell population in $Rora^{+/+}$ and $Rora^{+/-}$ from 1 to 24 months. Values are mean ± S.E.M.

Figure 4

Averaged laterolateral distribution of the Purkinje cell (PC) population at 1 and 24 months in $Rora^{+/-}$ (A) and $Rora^{+/+}$ (B) mice.

A: In $Rora^{+/-}$ mice no significant PC loss is observed between 1 and 24 months of age.

B: In $Rora^{+/+}$ mice there a significant PC loss between 1 and 24 months. This loss is evenly distributed along the laterolateral extent of the cerebellum. Values are mean ± S.E.M.

Figure 5

Histogram of the Purkinje cell population in both $Rora^{+/-}$ and $Rora^{+/sg}$ from 1 to 9 months. Values are mean ± S.E.M. Asterisks indicate that the differences between both genotypes are statistically significant. **P<0.001

|  | 1 month | 24 months |
|---|---|---|
| Rora+/+ | 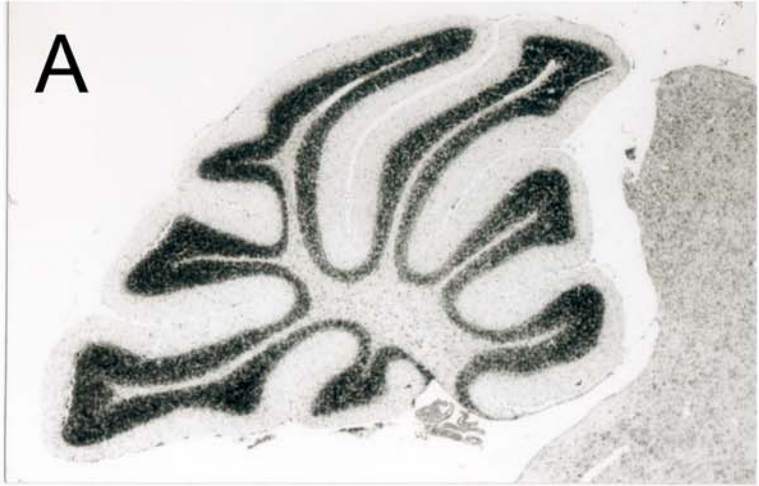 | 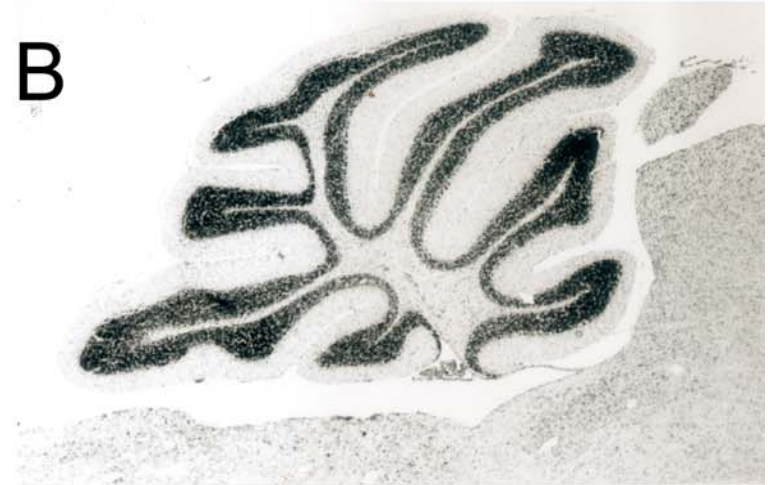 |
| Rora+/- | 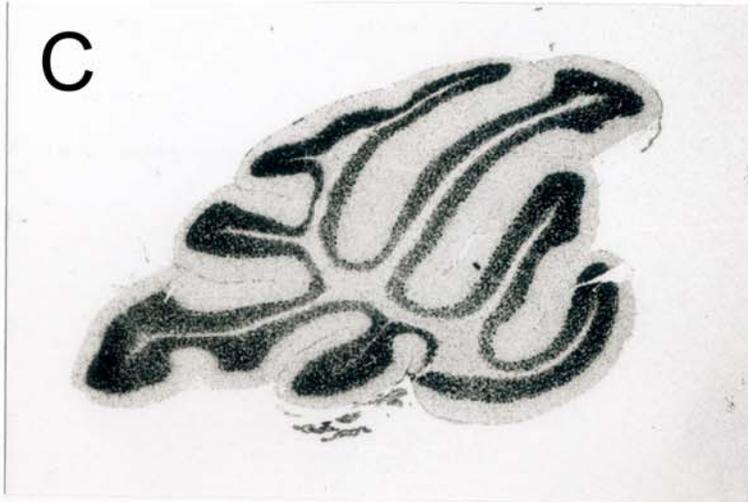 | 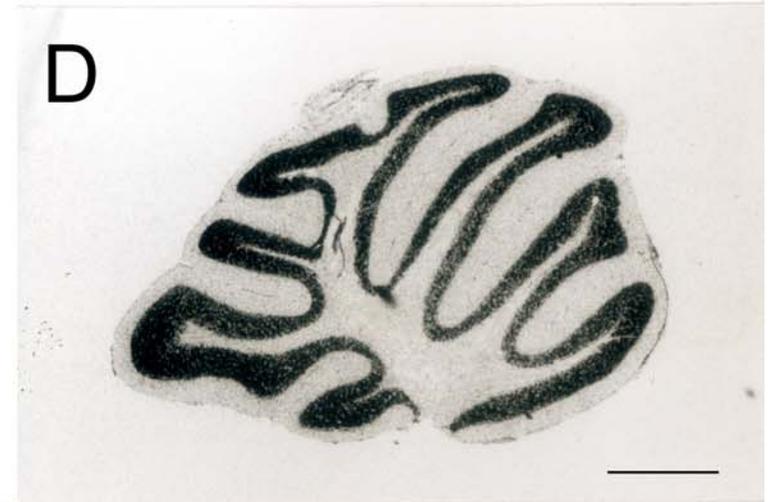 |

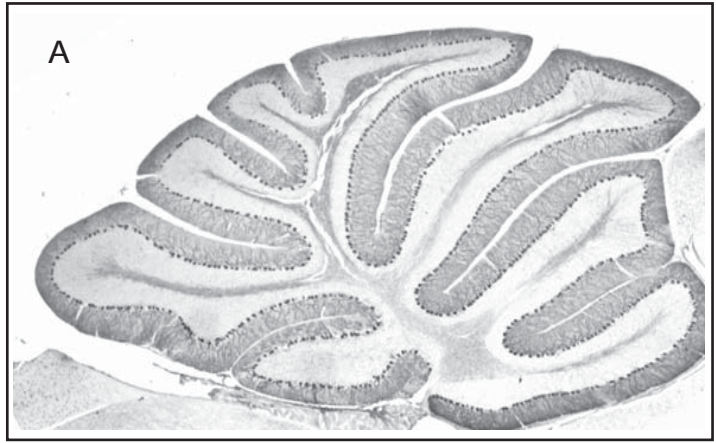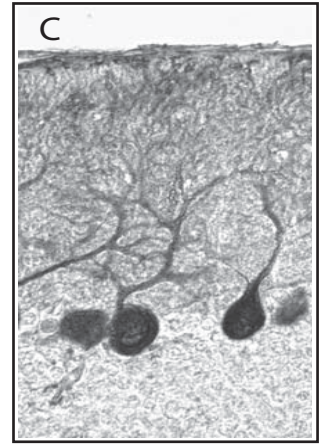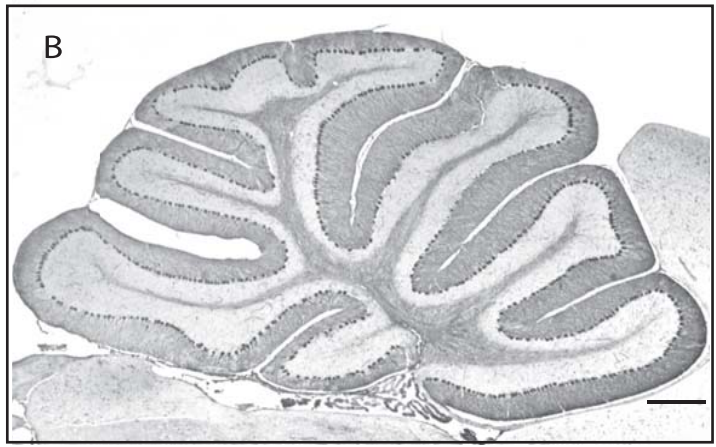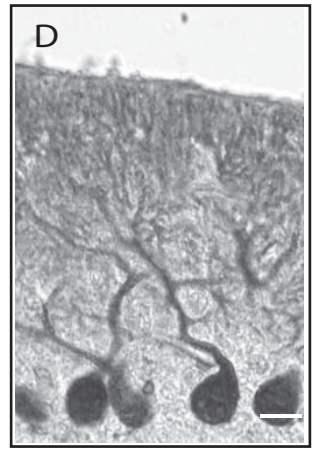

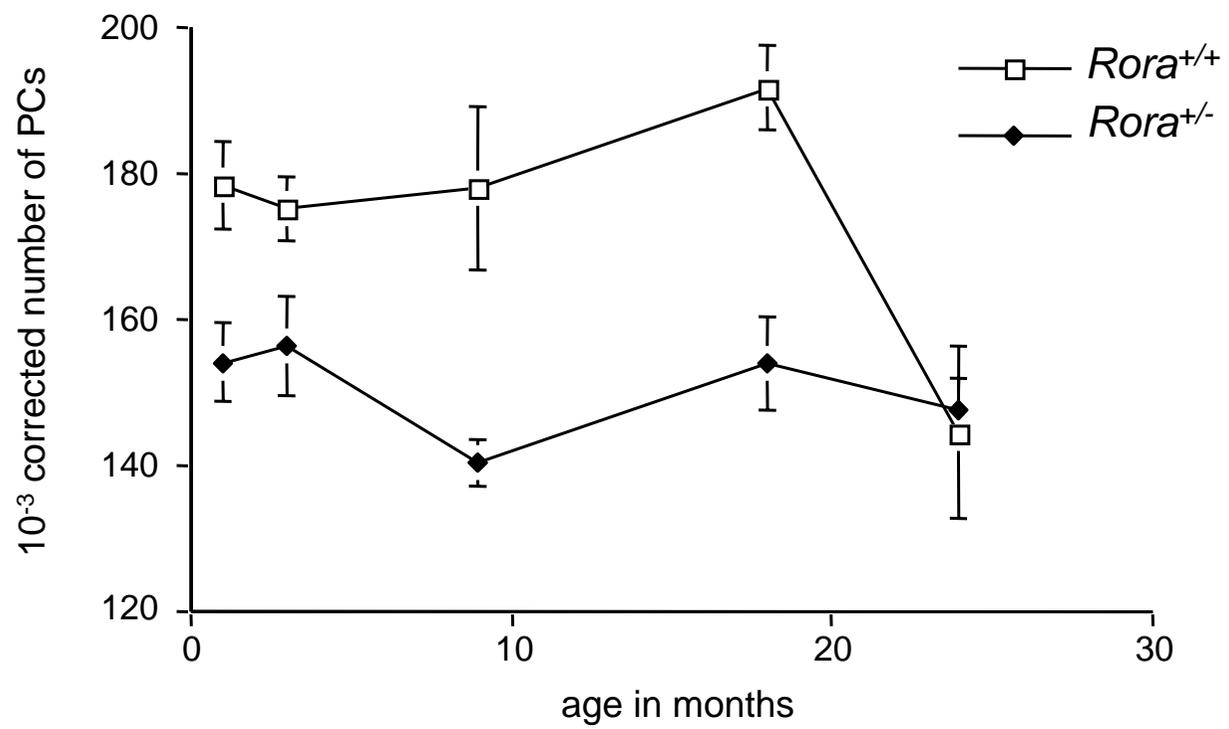

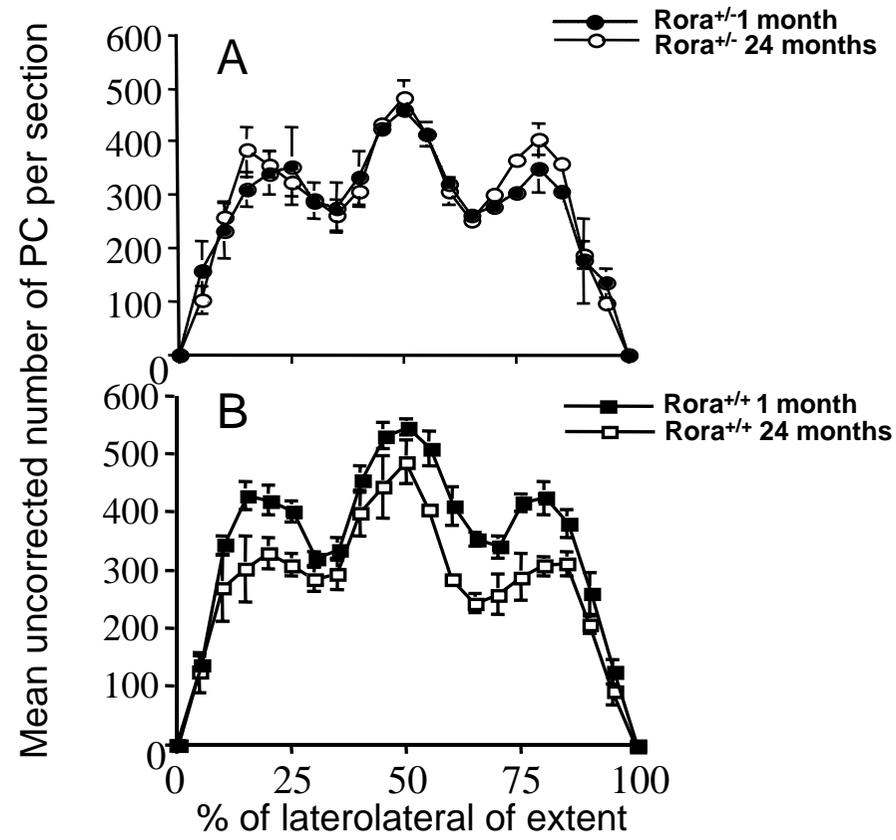

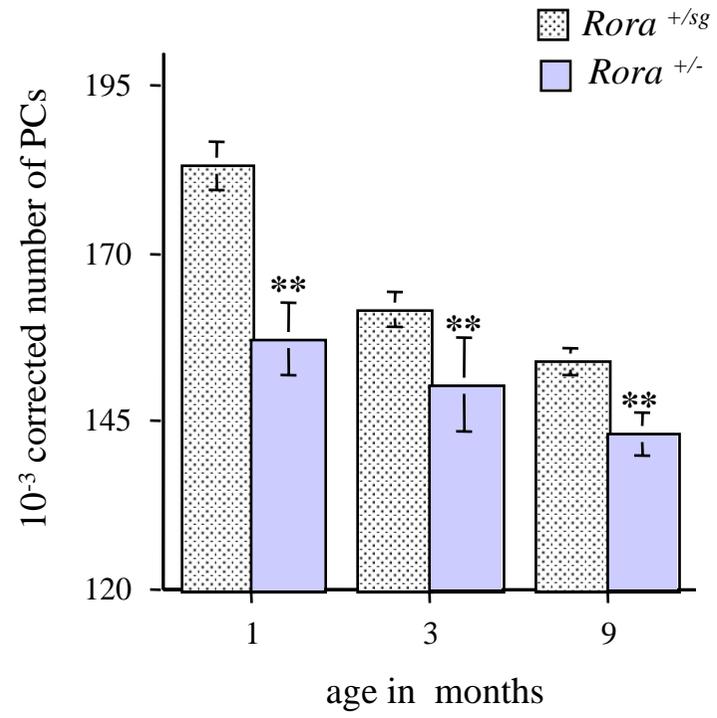